\documentclass{emulateapj}

\def\kms{\ifmmode{\rm km\thinspace s^{-1}}\else km\thinspace s$^{-1}$\fi}
\def\ms{\ifmmode{\rm m\thinspace s^{-1}}\else m\thinspace s$^{-1}$\fi}

\shortauthors{Torres et al.}
\shorttitle{Blend scenarios for OGLE-TR-56}

\begin{document}

\title{Testing blend scenarios for extrasolar transiting planet
candidates. II. --- OGLE-TR-56}

\author{Guillermo Torres\altaffilmark{1},
	Maciej Konacki\altaffilmark{2},
	Dimitar D.\ Sasselov\altaffilmark{1},
	Saurabh Jha\altaffilmark{3}
}

\altaffiltext{1}{Harvard-Smithsonian Center for Astrophysics, 60
Garden St., Cambridge, MA 02138, USA}

\altaffiltext{2}{California Institute of Technology, Div.\ of
Geological \& Planetary Sciences 150-21, Pasadena, CA 91125, USA}

\altaffiltext{3}{Department of Astronomy, University of California,
Berkeley, CA 94720, USA}

\email{gtorres@cfa.harvard.edu}

\begin{abstract} 

We re-examine the photometric and spectroscopic evidence available for
the star OGLE-TR-56, recently discovered to harbor a giant planet
presenting transits and orbiting with a period of 1.21 days. We
investigate the possibility that the observational signatures reported
might be the result of ``blending" with the light of an eclipsing
binary along the same line of sight. Using techniques developed
earlier we perform fits to the light curve under a variety of blend
configurations, subject to all observational constraints, and we infer
further properties of these possible blends. We then carry out
realistic end-to-end simulations based on those properties in order to
quantify the spectral line asymmetries and radial velocity variations
expected from such scenarios. We confront these calculations with the
observations.  The results from these models are clearly inconsistent
with the measured radial velocity and bisector span variations, ruling
out blends and confirming the planetary nature of the companion. The
example of OGLE-TR-56 serves to illustrate the sort of tests that can
and should be performed on transiting planet candidates to eliminate
the possibility of ``false positives", and in particular,
line-of-sight contamination from eclipsing binaries. 
	
\end{abstract}


\section{Introduction}
\label{sec:intro}

The search for extrasolar giant planets has entered an exciting new
era, one in which these objects can be discovered indirectly by the
tiny photometric signatures they produce as they transit across the
disk of their parent star (periodic $\sim$1\% drops in brightness).
Numerous photometric search programs are underway or under development
worldwide \citep[see][]{Horne:03}. Six stars with transiting planets
have been found so far: HD~209458 \citep{Henry:00, Charbonneau:00},
OGLE-TR-56 \citep{Konacki:03a}, OGLE-TR-113 \citep{Bouchy:04,
Konacki:04}, OGLE-TR-132 \citep{Bouchy:04}, TrES-1 \citep{Alonso:04},
and OGLE-TR-111 \citep{Pont:04}.\footnote{Another case, OGLE-TR-3, was
reported recently by \cite{Dreizler:03} and claimed to harbor a
transiting planet with a period of 1.19 days, but additional studies
showed that it is most likely a false positive \citep{Konacki:03b}.}
The OGLE-TR planets were originally detected as low-amplitude transit
events by the most successful of the photometric surveys, OGLE-III
\citep[Optical Gravitational Lensing Experiment;][]{Udalski:02b,
Udalski:02c, Udalski:03}, and later confirmed by radial-velocity
follow-up.  The planet around HD 209458 was originally discovered by
its radial velocity signature ---the small reflex motion of the star
in response to the pull from the planet--- and only later was it found
to undergo transits.  It is the Doppler technique, in fact, that has
been the most successful in discovering extrasolar planets, having
produced more than 100 giant planets to date \citep{Schneider:04}. 

Both the Doppler and the transit search methods suffer from the
problem of astrophysical ``false positives", which refers to other
phenomena that can produce the same observational signature but do not
involve planets.  In the case of the Doppler searches the concern is
that similar velocity variations might result from chromospheric
activity on the star (e.g., spots), sub-surface convection
(granulation), other line asymmetries from undetected stellar
companions, or even stellar oscillations \citep[see, e.g.,][]{Gray:97,
GrayH:97, Gray:98}. For the transit searches the main contaminants are
(1) eclipsing binaries with grazing geometry, (2) transits of a small
star in front of a large star (such as an M dwarf passing in front of
an F star, or a solar-type star in front of a giant), and (3)
eclipsing binary systems with deep eclipses, blended with the light of
another star along the same line of sight that dilutes the minima.  In
the latter case (referred to as a ``blend"), the intruding star may or
may not be physically associated with the binary. It is these blend
scenarios that are the subject of this paper. 

Cases that only involve an eclipsing binary with no contaminants
(scenarios [1] and [2] above) are relatively easy to dismiss with
radial velocity measurements since the expected amplitudes are
typically several tens of~\kms, as opposed to a few hundred~\ms\ for
the case of a Jupiter-mass planet around a solar-type star. Blend
configurations, on the other hand, can be much more difficult to rule
out as the main star may show little or no velocity variation at all,
and may therefore appear to have a very low-mass companion in orbit
when in reality it does not.  This is particularly the case for very
faint candidates ($V = 14$--19) such as those being reported from
surveys such as OGLE-III \citep[e.g.,][]{Udalski:02a}, EXPLORE
\citep{Mallen-Ornelas:03}, MACHO \citep{Drake:04}, and others.  For
these objects the precision of the velocities is much lower than for
the typically brighter stars ($V < 11$) that have been targeted in
Doppler searches, and the potential for confusion is greater because
they also tend to be in very crowded fields. It is therefore important
to examine these cases very carefully. 

In a recent paper \citep[][hereafter Paper~I]{Torres:04b} we described
a procedure to model the light curves of transit candidates as the
combination of three stars forming the blend, and we applied it to the
case of OGLE-TR-33, a candidate that turned out to be a false
positive. We showed how such modeling tools allow one to predict other
observable properties of the blend and can become a useful guide to
look for additional signs of a contaminating eclipsing binary. One
such sign that can easily be overlooked is asymmetries in the spectral
lines, which can lead to spurious velocity measurements.  In the
present paper we develop procedures to simulate line asymmetries and
their effects on the measured velocities of a transit candidate
resulting from typical blend scenarios.  We illustrate these
techniques by carrying out a critical examination of the case of
OGLE-TR-56 ($V = 16.6$), the first extrasolar transiting planet to
emerge from the lists of very faint candidates, and the one with the
shortest orbital period so far (1.21~days). We perform blend
simulations subject to all observational constraints in an attempt to
explain the system as a false positive, but despite all our efforts we
are unable to find a plausible configuration involving a contaminating
eclipsing binary that does not violate one or another of those
constraints by a large margin. This lends further support to the
planetary nature of the companion of OGLE-TR-56. 
	
\section{OGLE-TR-56: The spectroscopic evidence}
 \label{sec:ogle56}

The original detection of periodic transit-like events in OGLE-TR-56
was reported by \cite{Udalski:02b}. The star was part of the second
batch of candidates produced by the OGLE-III photometric survey, which
targeted three fields toward the Galactic center.  High-resolution
spectroscopic follow-up observations were conducted by our team in
July 2002, and small but significant radial velocity variations were
announced by \cite{Konacki:03a} on the basis of measurements on three
nights near the quadratures, implying the presence of a planetary-mass
object in orbit. As the first such claim among all the transit
candidates reported until then ---and by far the faintest candidate
ever considered to have a chance of harboring a planet--- those
results were carefully scrutinized by the community and were received
quite understandably with a measure of skepticism.  The initial
concerns on the observational side regarding this particular detection
focussed on two issues: (a) The \emph{reality} (or statistical
significance) of the radial velocity variations, especially since the
claim was based on only 3 averaged measurements; and (b) the
\emph{interpretation} of the velocity variations, given the crowding
in the field and the increased probability of a blend.\footnote{Two
additional concerns have to do with (c) whether a planet in such a
tight orbit such as OGLE-TR-56 can exist at all \citep[which theory
appears not to rule out; see][]{Trilling:98, Baraffe:04}, and (d) why
Doppler searches among brighter stars have not found additional
examples in very short-period orbits ($\leq 2.5$~days) despite being
sensitive to them. The latter may perhaps be the result of a much
lower frequency of occurrence of these ``very hot Jupiters", coupled
with very strong selection toward short periods in transit surveys
such as OGLE \citep[see][]{Bouchy:04, Gaudi:04}.}

Such concerns are likely to be repeated for other faint candidates
that might be confirmed in the future, and it is therefore necessary
to consider them very seriously. The first concern is related mostly
to errors of measurement (accidental or systematic), and whether the
results might be due, for example, to limitations in the reduction
and/or analysis techniques or to the stability of the spectrograph
employed \citep[in this case, HIRES on the Keck~I
telescope;][]{Vogt:94}. For OGLE-TR-56 these issues were addressed by
\cite{Konacki:03b}, who showed that systematic errors in the
velocities are no more than about 100~\ms\ and accidental errors are
somewhat smaller, making the reported peak-to-peak velocity change
(initially estimated to be $\sim 340~\ms$) statistically significant
considering that the phasing in the orbit is known and fixed from the
photometry.  Eight additional measurements with the same instrument
were reported more recently by \cite{Torres:04a}, and show even more
clearly that the velocity is changing as expected from the phasing in
the orbit. The revised amplitude folding together all the measurements
is larger than originally estimated, but it is now much better
characterized (peak-to-peak variation near 600~\ms).  More
importantly, independent confirmation of velocity variations has been
reported by another group using different instrumentation.
\cite{Mayor:03} obtained five measurements of OGLE-TR-56 during the
commissioning of the new HARPS instrument on the ESO 3.6-m telescope
in June of 2003, achieving a nominal precision of about 30~\ms.  Their
preliminary analysis gives much the same amplitude as
\cite{Torres:04a}. 

With the reality of the velocity variations now beyond doubt, the
issue of their interpretation becomes the main concern for OGLE-TR-56,
and is the motivation for the remainder of this paper.  Can the
observations be explained by a blend scenario, with an eclipsing
binary along the same line of sight contaminating the light of the
main star?  If so, the velocity variations measured would be caused by
blending with the spectral lines from the primary of the eclipsing
binary moving back and forth relative to the main star, with a period
of 1.21~days. Are these extra lines visible in the spectra?  Strong
and variable line asymmetries are also to be expected from such a
scenario. Are the measured asymmetries \citep{Torres:04a} consistent
with a blend model?  Given the importance of OGLE-TR-56 as the first
example of a seemingly new class of planetary objects under extreme
conditions of proximity to the parent star, the answers to the
questions above must be \emph{quantitative} rather than
\emph{qualitative}. We begin in the next section by modeling the light
curve in detail following the procedures we introduced in Paper~I for
OGLE-TR-33. 

\section{Light curve fits to OGLE-TR-56 under a blend scenario}
 \label{sec:lightcurve}

The shape of a light curve showing transit-like events contains
valuable information that can be used in some cases to recognize false
positive scenarios.  For example, \cite{Drake:03} and \cite{Sirko:03}
showed that the variations outside of eclipse, as measured from a
Fourier analysis, can often indicate that the companion is stellar as
opposed to substellar, from effects due to tidal distortions,
reflection, or gravity brightening. An isolated eclipsing binary
(e.g., one with grazing geometry, or consisting of a small star
passing in front of a much larger star) can be ruled out as the
explanation for the light curve of OGLE-TR-56 because of the very
small velocity amplitude measured. However, an eclipsing binary
blended with a another star remains a possibility, in principle, and
some of its properties could still show through in the light curve,
even though they would be diluted. But as reported by the authors
above, the Fourier decomposition test for OGLE-TR-56 comes out
negative, indicating no significant curvature outside of eclipse. 

A different kind of test was proposed by \cite{Seager:03} based on the
morphology of the transits themselves. They showed that the mean
density of the primary star, as well as the radius of the companion,
can be estimated by careful measurement of the depth of the transit
(relative change in flux, $\Delta F$), its total duration ($t_T$), and
the duration of the flat portion ($t_F$), along with the knowledge of
the orbital period. Any inconsistency in the mean density with values
for normal (main-sequence) stars, or with additional information
available for the primary, would be an indication of a blend. Once
again this test comes out negative for OGLE-TR-56. We measure $\Delta
F = 0.016$, $t_T = 0.060$, and $t_F = 0.036$ (the latter two in phase
units), which when combined with the orbital period of 1.2119189~days
\citep{Torres:04a} lead to a primary star very similar to the Sun in
mean density, as well as in both mass and radius when a typical
mass-radius relation for main sequence stars is adopted. Indeed the
temperature estimated for OGLE-TR-56 from our high-resolution spectra
is solar-like \citep[$\sim$5900~K;][]{Sasselov:03}. 

More detailed information on the properties of a possible blend can be
obtained from a direct fit to the light curve as described in Paper~I.
Briefly, the photometry is modeled using the eclipsing binary computer
code EBOP \citep[see][]{Nelson:72, Etzel:81, Popper:81}, fully
accounting for effects such as stellar oblateness, reflection, limb
darkening, and gravity brightening.  The properties of the three stars
composing the blend (which are assumed to be on the main sequence) are
parameterized in terms of their mass and are constrained to fit
suitable model isochrones. These isochrones may be different for the
binary and for the main star if they do not form a physical triple
system. In that case, differential extinction must also be accounted
for since they can be at different distances from the observer. All
other relevant properties of the stars such as their radii and
luminosities are taken also from the isochrones (see Paper~I for
details).  The important advantage of this procedure is that it allows
testable predictions for some of the other properties of the eclipsing
binary that are potentially observable. These properties include the
brightness of the primary in the eclipsing binary relative to the main
star, its velocity semi-amplitude, and its projected rotational
velocity (assuming its spin is synchronized with the orbital motion).
This was illustrated in Paper~I with the example of the blend case
OGLE-TR-33.  The same procedure is applied here to OGLE-TR-56. 

As in Paper~I we make use of the model isochrones by \cite{Girardi:00}
to describe the properties of the stars in the blend\footnote{For
low-mass stars these models become increasingly unrealistic partly
because of shortcomings in the treatment of convection, in the
opacities, in the equation of state, and the use of gray boundary
conditions, all of which become very important in the lower main
sequence.  In particular, theoretical radii from these and all current
stellar evolution models are known to be underestimated by 10\% or
more compared to measurements of well-studied low-mass eclipsing
binaries \citep[see, e.g.,][]{Torres:02, Ribas:03}. While other models
for cool stars are somewhat more sophisticated
\citep[e.g.,][]{Siess:97, Baraffe:98}, our reason for choosing this
particular set of isochrones is that they span the largest range of
masses for our application, beginning at 0.15~M$_{\sun}$. Despite
this, slight extrapolations to even lower masses were necessary in
some cases.  Given the importance of the model radii in our
simulations for determining key properties of the eclipses such as
their depth and duration, we have attempted to bring them into closer
agreement with the radii of real stars by applying ad-hoc corrections
based on a careful comparison with accurately measured radii for
late-type stars.  The correction factors are mass-dependent, and
average 1.1 over the mass range of interest for our work.}.  The
photometric data for OGLE-TR-56 are described by \cite{Torres:04a},
and we adopt the measurement errors reported by the OGLE
team\footnote{See {\tt
http://bulge.princeton.edu/$\sim$ogle/ogle3/transits/ ogle56.html}, and
\cite{Kruszewski:03}.\label{foot:1}}. Those observations were made in
the $I$ band. We begin by assuming the eclipsing binary and the third
star are physically associated, and we therefore use the same
isochrone for both with an adopted age of 3~Gyr \citep{Sasselov:03}
and solar metallicity ($Z = 0.019$). We refer to this model as {\tt
MODEL1}. We follow the notation of Paper~I and refer to the primary
and secondary in the eclipsing binary as star~1 and star~2,
respectively, while the main star that dilutes the eclipses of the
binary is referred to as star~3.  We assign to star~3 a mass on the
isochrone such that the corresponding temperature agrees with the
spectroscopic determination for OGLE-TR-56 ($M_3 = 1.05$~M$_{\sun}$,
$T_{\rm eff} = 5900$~K). The inclination angle is assumed to be
90\arcdeg\ for simplicity.

\begin{figure}[!ht]
\vskip -0.1in
\hskip 0.2in\includegraphics[angle=0,scale=0.4]{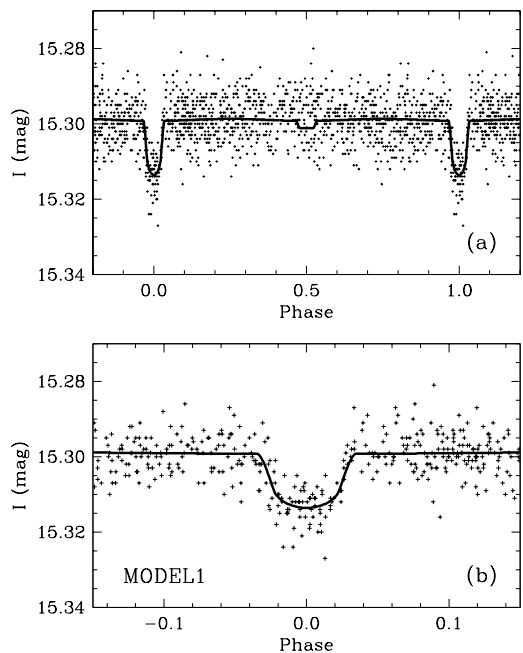}
\vskip 0.3in
 \figcaption[Torres.fig01.ps]{(a) Blend model fit to the light curve of
OGLE-TR-56 ({\tt MODEL1}). The main star (star~3) and the
contaminating eclipsing binary are assumed to be physically associated
(age = 3~Gyr). The masses of the stars in the eclipsing binary (star~1
and star~2) are $M_1 = 0.89$~M$_{\sun}$ and $M_2 = 0.13$~M$_{\sun}$,
and star~3 contributes 71\% of the light in the $I$ band.  The
predicted light ratio in the visual band is $L_1/L_3 = 0.35$. (b)
Enlargement of the primary eclipse.\label{fig:model1}}
 \end{figure}

Our blend model adjusted to the light curve is shown in
Figure~\ref{fig:model1}. The resulting properties of the stars in the
eclipsing binary are $M_1 = 0.89$~M$_{\sun}$ (approximately spectral
type \ion{G8}{5}) and $M_2 = 0.13$~M$_{\sun}$ (\ion{M4-5}{5}), and the
inferred distance to the system is $\sim$1.8~kpc. Very shallow
secondary eclipses only 0.002~mag deep are also predicted by this
model, but they are undetectable given the precision of the
observations (typical errors of $\sim$0.006~mag for an individual
measurement). We note that the fit to the primary eclipse with this
blend model is essentially indistinguishable from a fit that models a
true planetary transit \citep[see][Figure~4]{Torres:04a}. Thus, the
shape of the light curve alone is insufficient to discriminate between
the two. The fitting procedure provides all the physical properties of
star~1 from the isochrone, so that it is possible to estimate the
semi-amplitude of its radial-velocity orbit ($K_1 = 26$~\kms) and also
its projected rotational velocity ($v_1 \sin i = 36$~\kms), under the
plausible assumption that it is synchronized with the orbital motion
(given the short orbital period). The brightness of star~1 relative to
the main star is predicted to be $L_1/L_3 = 0.35$ in the visual
band\footnote{The symbol $L$ is used throughout this paper to denote
brightness (``light").}. The latter quantity conflicts with the
spectroscopic observations for OGLE-TR-56, which place an upper limit
on the presence of lines from another star in the spectrum of about
3\% of the light of the main star \citep{Konacki:03b}. A star as
bright as 35\% of the light of star~3 would be fairly obvious in the
spectra, despite the rotational broadening.  We conclude that this
particular blend scenario is inconsistent with the data\footnote{Tests
show that adjusting the age of the isochrone does not qualitatively
change the conclusion. With increasing age $L_1/L_3$ decreases
somewhat as star~3 evolves upward in luminosity, but not enough before
a star with a mass appropriate for the temperature of OGLE-TR-56
leaves the main sequence altogether.  The latter occurs for ages
greater than about 7~Gyr, at which $L_1/L_3$ has only decreased to
approximately 0.20.  Further tests with lower inclination angles for
the binary show that the $L_1/L_3$ \emph{increases} as the system
departs from an edge-on orientation, making the disagreement with the
observations even worse.}.  Additional difficulties with {\tt MODEL1}
arise from the magnitude of the line asymmetries that are expected. We
defer the discussion of these issues to \S\ref{sec:asymmetries}. 

\begin{figure}[!hb]
\vskip -0.3in
\includegraphics[angle=0,scale=0.43]{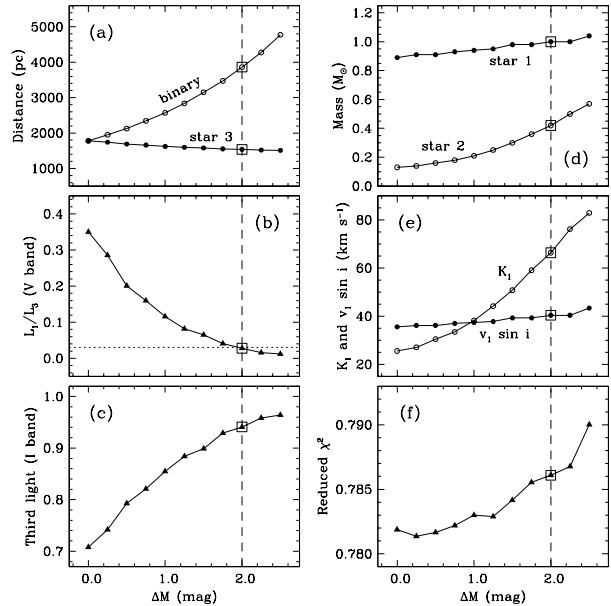}
\vskip 0.1in
 \figcaption[Torres.fig02.ps]{Results from a grid of light curve fits
to the photometry of OGLE-TR-56, with the eclipsing binary placed at
increasing distances behind the main star (star~3), as parameterized
by the difference in the distance modulus $\Delta M$.  The fit marked
with squares is referred to in the text and in Table~1
as {\tt MODEL2}. The leftmost points correspond to {\tt MODEL1}. (a)
Fitted distance in parsecs to the binary and to star~3; (b) Light
ratio $L_1/L_3$ in the visual band.  The horizontal dotted line
represents the upper limit estimated from the high-resolution spectra
($\sim$3\%); (c) Contribution of star~3 to the total light in the $I$
band (``third light" in eclipsing binary terminology); (d) Best fit
masses for the primary and secondary in the eclipsing binary; (e)
Predicted radial velocity semi-amplitude and projected rotational
velocity for the primary in the eclipsing binary (star~1); (f) Reduced
$\chi^2$ for the solution (that the values are lower than unity
suggests the photometric errors are slightly
overestimated).\label{fig:distmod}}
 \end{figure}

Since the main problem with {\tt MODEL1} seems to be the excessive
brightness of star~1, we explored ways of making star~1 fainter while
still providing a reasonably good fit to the light curve. One such way
is to remove the requirement that the eclipsing binary and the main
star be in a physical triple, and to allow the binary to be in the
background.  In this case the age of the isochrone for the binary can
in principle be different.  For lack of a better estimate we retain
the 3~Gyr isochrone for both, but the binary is no longer constrained
to be at the same distance as star~3. We performed extensive modeling
of the photometry for a range of distances between the binary and
star~3 parameterized, for convenience, in terms of the difference in
distance modulus $\Delta M$ (see Paper~I). We assumed an edge-on
orientation, as before.  The results are summarized in
Figure~\ref{fig:distmod}, where we show how the parameters of star~1
and star~2 change as the binary is placed farther behind star~3. As
expected the observed light ratio $L_1/L_3$ decreases with distance
(Figure~\ref{fig:distmod}b), and reaches the upper limit determined
spectroscopically (0.03) at $\Delta M \approx 2.0$ (binary
approximately 2.3~kpc behind star~3; Figure~\ref{fig:distmod}a).  We
refer to this solution as {\tt MODEL2}, and indicate it with squares
in the figure.  The mass of star~1 that provides the best fit has
increased at this distance by about 12\% compared to {\tt MODEL1},
while that of star~2 is more than a factor of 3 larger than before
(Figure~\ref{fig:distmod}d). The reason for these changes can be
understood as follows. As the binary is pushed back and its
contribution to the total light decreases, the eclipse becomes more
diluted (shallower), and so in order to maintain the observed depth
the relative size of the secondary in the binary (star~2) must
increase relative to the primary (star~1), which implies increasing
the mass of star~2.  This, in turn, will affect the total duration of
the eclipse for a fixed star~1, both from the change in the relative
size of star~2 and star~1, and from the change in semimajor axis for a
fixed orbital period \citep[see][]{Seager:03}. In order to conform to
the observed transit duration, the mass and size of star~1 then need
to increase slightly to compensate.  Accordingly, the predicted $v
\sin i$ of star~1 shows only a small dependence with $\Delta M$, while
$K_1$ varies significantly as the mass of the companion increases. The
quality of the fits degrades as the binary is placed farther in the
background (Figure~\ref{fig:distmod}f), but remains more or less
acceptable at $\Delta M = 2.0$. That fit is shown later in
Figure~\ref{fig:model23}a along with the light curve from {\tt MODEL1}
for comparison. The consequences of {\tt MODEL2} for the line
asymmetries and expected velocity variations are explored in the next
section. 

\begin{figure}[!ht]
\vskip -0.4in 
\hskip -0.3in\includegraphics[angle=0,scale=0.50]{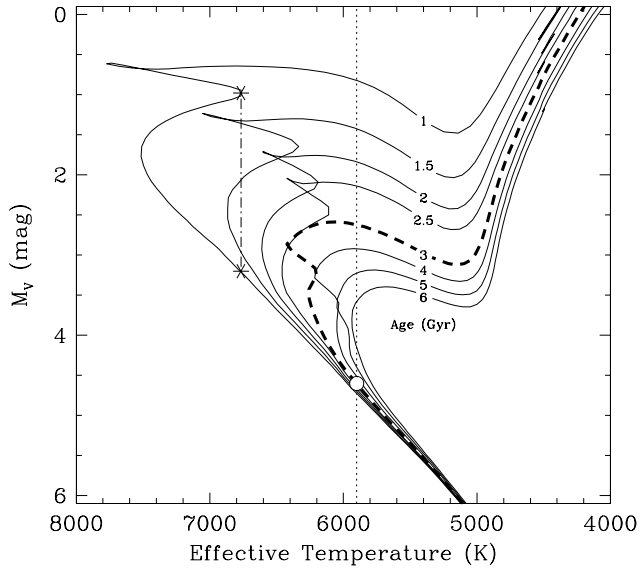} 
\vskip -0.2in
 \figcaption[Torres.fig03.ps]{Isochrones from the Yonsei-Yale series
of evolutionary models \citep{Yi:03} for a range of ages labeled in
Gyr.  The 3~Gyr isochrone for OGLE-TR-56 is represented with a heavy
dashed line.  As an example, the asterisks show two possible locations
of a main-sequence star at a fixed age with the same effective
temperature, but with brightness differing by about 2 mag. The effect
of choosing the brighter location for star~3 in a blend model would be
to decrease $L_1/L_3$. The open circle on the 3-Gyr isochrone
corresponds to the position of OGLE-TR-56 on the main sequence, at the
measured temperature of 5900~K (vertical dotted line). At these low
temperatures the flexibility of choosing the brighter of two
vertically aligned locations for star~3 with significantly different
brightness is effectively eliminated because of the change in the
shape of the isochrones.\label{fig:isochrones}}
 \end{figure}

If we wish to keep the three stars at the same distance, as in a
triple system, an alternate way of decreasing the relative brightness
of star~1 to comply with the spectroscopic constraint is to make
star~3 itself intrinsically \emph{brighter}, e.g., by evolving it from
the zero-age main sequence. There are good reasons for assuming
physical association that have to do with the apparent symmetry of the
velocity variations (see \S\ref{sec:asymmetries}).  The brightness
change required here is roughly a factor of 10 (about 2.5 magnitudes)
in order to reduce the brightness ratio $L_1/L_3$ from 0.35 ({\tt
MODEL1}) to the observational upper limit of 0.03 \citep{Konacki:03b}.
A similar situation was described in Paper~I for the case of
OGLE-TR-33, where it was shown that by fine-tuning the age an
isochrone could be found in which a slightly more evolved star~3 could
be about 2 magnitudes brighter at the same effective temperature
(constrained by the spectroscopy).  The location of star~3 in the H-R
diagram would be beyond the turnoff but still on the main sequence,
near the point of hydrogen exhaustion \citep[see Figure~5
of][]{Torres:04b}.  OGLE-TR-56 is cooler than OGLE-TR-33, however, and
a glance at the morphology of the isochrones reveals that a similar
argument does not work for a temperature of 5900~K, at any age. This
is illustrated in Figure~\ref{fig:isochrones}, where instead of the
\cite{Girardi:00} models we have used those of \cite{Yi:03} for their
finer resolution near the turnoff. It is seen, for example, that the
``blue hook" feature gradually disappears for older ages, reducing the
difference in brightness between its red end (coolest and most
luminous point on the main sequence beyond the turnoff) and a location
below it with the same temperature. The change in shape as the age
increases coincides with the disappearance of convective cores at
masses slightly larger than the Sun. Unless one is willing to accept
that OGLE-TR-56 is in the rapid phase of evolution often referred to
as the Hertzprung gap (horizontal portions of the isochrones where
stars are evolving mostly toward cooler temperatures without changing
their brightness significantly), the brightness of star~3 cannot be
much larger than indicated by the open circle, at its location on the
main sequence. 

\begin{figure}[!ht]
\vskip -0.1in
\includegraphics[angle=0,scale=0.43]{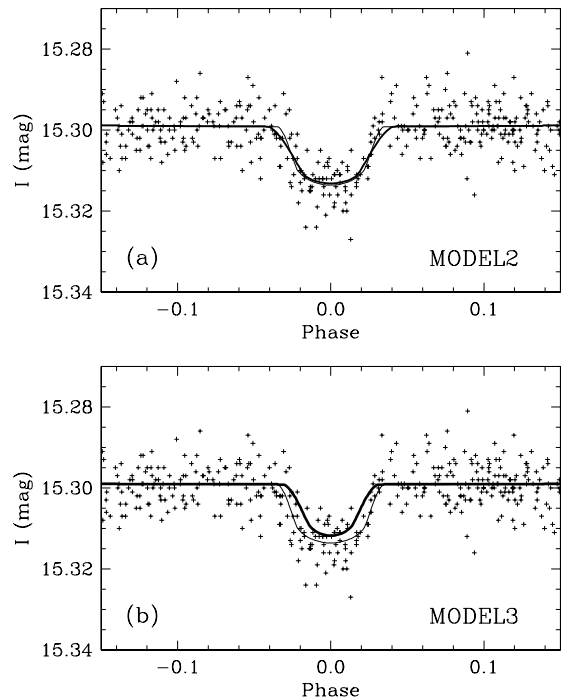}
\vskip 0.4in
 \figcaption[Torres.fig04.ps]{Two blend models for OGLE-TR-56 (thick
lines) that satisfy the spectroscopic constraint on the brightness of
star~1 relative to star~3 in the $V$ band ($\leq 3$\%). The fit for
{\tt MODEL1} from Figure~\ref{fig:model1}b is superimposed in each
panel for reference (thin line). (a) {\tt MODEL2}, in which the binary
is in the background of star~3 ($\sim$2.3~kpc behind, in this case;
see Figure~\ref{fig:distmod}). (b) {\tt MODEL3}, in which the binary
and star~3 are at the same distance (as in a physical triple), but
star~1 is constrained to have a brightness of 3\% relative to the main
star.  \label{fig:model23}}
 \end{figure}

For completeness we explored also a blend scenario in which the three
stars form a physical triple system, but star~1 is constrained to have
a relative brightness in the $V$ band consistent with the upper limit
from spectroscopy ($\sim$3\%).  Only the properties of star~2 were
adjusted, and we refer to this as {\tt MODEL3}. The inclination angle
was again fixed at 90\arcdeg, and the adopted age is the same as
before (3~Gyr), although this does not qualitatively change the
results.  This solution is noticeably worse than the previous two
(reduced $\chi^2$ about 8\% larger), but at least it still reproduces
the total duration and depth of the eclipse tolerably well (see
Figure~\ref{fig:model23}b).  We summarize the properties of this and
the other two models in Table~1. 

\begin{deluxetable}{lccc}
\tablenum{1}
\tablehead{
\colhead{~~~~~~~~~~Parameter~~~~~~~~~~~~~} & \colhead{MODEL1} &
\colhead{MODEL2} & \colhead{MODEL3}}
\tablecaption{Summary of blend models for OGLE-TR-56.}
\startdata
$M_1$ (M$_{\sun}$)\dotfill                          & 0.89    & 1.00    & 0.60    \\
$M_2$ (M$_{\sun}$)\dotfill                          & 0.13    & 0.42    & 0.22    \\
$L_1/L_3$ ($V$ band)\dotfill                   & 0.35    & 0.03    & 0.03    \\
$L_3/L_{\rm tot}$ ($I$ band) (\%)\tablenotemark{a}\dotfill & 71      & 94      & 94      \\
$K_1$ (\kms)\dotfill                                & 26      & 66      & 50      \\
$v_1 \sin i$ (\kms)\dotfill                         & 36      & 40      & 26      \\
Distance to star~3 (kpc)\dotfill                    & 1.78    & 1.54    & 1.54    \\
Distance to binary (kpc)\dotfill                    & 1.78    & 3.86    & 1.54    \\
$\Delta M$ (mag)\tablenotemark{b}\dotfill           & 0.00    & 2.00    & 0.00    \\
$a_v$ (mag~kpc$^{-1}$)\tablenotemark{c}\dotfill     & 0.56    & 0.66    & 0.66    \\
Reduced $\chi^2$ of the fit\tablenotemark{d}\dotfill & 0.78188 & 0.78609 & 0.84734 \\
\enddata
 \tablenotetext{a}{Relative contribution of star~3 to the total light
of the system.}
 \tablenotetext{b}{Difference in the distance modulus between the
binary and star~3.}
 \tablenotetext{c}{Coefficient of differential extinction in the $V$
band, adjusted to reproduce the observed system magnitudes in the $V$
and $I$ bands (see Paper~I).}
 \tablenotetext{d}{The values significantly lower than unity imply
that the measurement errors are slightly overestimated, by roughly
10\%.}
 \end{deluxetable}

As in the case of the first scenario, {\tt MODEL2} and {\tt MODEL3}
predict shallow secondary eclipses with depths of $\sim$0.002~mag and
$\sim$0.006~mag, respectively, which would be difficult to see. 
	
\section{Simulating line asymmetries and radial velocity variations}
\label{sec:asymmetries}

In blend scenarios such as those modeled in the previous section the
presence of lines from another star in the spectrum of OGLE-TR-56 will
affect the measurement of the radial velocities to some degree. As
mentioned above this is due to asymmetries introduced in the profiles
of the main star. The extent to which these asymmetries will distort
the radial velocities depends on the brightness of the secondary lines
and their relative displacement.  Intrinsic broadening may also be a
factor. Quantitative measures of the asymmetries for planet candidates
showing velocity variations are now often reported in the literature,
and in many cases they have turned out to be small. We believe,
however, that this alone is not sufficient evidence against a blend.
In this section we explore the magnitude of these effects by means of
simulations to see whether they can explain the measurements reported
for OGLE-TR-56 by \cite{Torres:04a}.

A similar philosophy was adopted by \cite{Santos:02} to investigate
small radial velocity variations and line asymmetries found in the
case of the star HD~41004. This is a close visual binary initially
thought to have a planetary companion around the primary star based on
high-precision Doppler monitoring performed by those authors. The
visual binary with an angular separation of 0\thinspace\farcs5 was
unresolved at the spectrograph, effectively making it a blend. Unlike
the case of OGLE-TR-56, however, there is no transit signature in
HD~41004 and therefore there is less information about the stars and
the geometry of the system.  \cite{Santos:02} approximated the
cross-correlation functions of the two stars in HD~41004 with
Gaussians, and sought to reproduce the measured velocity and bisector
variations by combining two Gaussians with appropriate parameters to
simulate the measured correlation functions.  From numerical
experiments they were able to infer the key parameters of the system,
and to fully explain the observations as the result of a possible
brown dwarf orbiting the visual companion instead of a planet around
the primary star.  Their results for the physical and orbital
properties of the visual companion were subsequently corroborated by
\cite{Zucker:03}. 

In this paper the situation is somewhat different in that our blend
models fitted to the light curve of OGLE-TR-56 constrain many of the
key properties of the stars. Our goal is then to see if these inferred
properties lead to a consistent picture of velocity and asymmetry
variations matching what is observed.  In addition, we wish to be as
realistic as possible in approximating the observations.  Therefore,
instead of adopting Gaussian profiles for the correlation functions,
we simulate composite spectra guided by the models in
\S\ref{sec:lightcurve}, by combining together calculated spectra for
star~1 and star~3 that conform to the stellar and orbital
characteristics implied by each blend scenario. We then measure radial
velocities by cross-correlation and estimate spectral line asymmetries
in these artificial spectra in exactly the same way as was done for
the original observations of OGLE-TR-56. 

For star~3 the calculated spectrum we used is the same one adopted by
\cite{Torres:04a}, which was derived from a fit to the observed
spectrum of OGLE-TR-56, and was used in that work as the template for
the cross-correlations. This calculated spectrum has an effective
temperature of 6000~K, surface gravity $\log g = 4.5$, and projected
rotational velocity $v \sin i = 3$~\kms. For a given blend model
star~1 is assigned the temperature and surface gravity inferred from
the isochrone for its mass, and a rotational velocity given by the
predicted value of $v_1 \sin i$. Star~1 was then scaled down by the
predicted light ratio $L_1/L_3$, and Doppler-shifted before adding it
to the spectrum of the main star.  The shift between star~1 and star~3
is controlled by the velocity amplitude of star~1 in its orbit ($K_1$)
and the difference $\Delta\gamma$ between its center-of-mass velocity
and the (presumably constant) velocity of star~3. Since the period and
epoch of transit are well known, the Doppler shift can be computed for
any instant of time.  $\Delta\gamma$ is a free parameter in principle,
although for scenarios such as {\tt MODEL1} and {\tt MODEL3} it is
expected to be small compared to $K_1$ given that the configuration is
assumed to be a hierarchical triple system. 

In an effort to approximate the real observations of OGLE-TR-56 as
closely as possible, the artificial composite spectra were divided
into sections corresponding precisely to each of the echelle orders of
the original observations \citep[see][]{Konacki:03b}, and rebinned to
exactly the same number of pixels as in the real data prior to further
analysis. The calculated spectra were initially generated at much
higher spectral resolution than the real observations (by a factor of
$\sim$10), and then broadened with the average instrumental profile of
the Keck spectrograph, which was assumed to be Gaussian in shape and
with FWHM $= 4.6$~\kms. 

Radial velocities from these artificial composite spectra were
measured by cross-correla\-tion using the IRAF\footnote{IRAF is
distributed by the National Optical Astronomy Observatories, which is
operated by the Association of Universities for Research in Astronomy,
Inc., under contract with the National Science Foundation.} task XCSAO
\citep{Kurtz:98}, combining the results for all orders.  The template
used is the same as for the real observations \citep{Torres:04a}, and
it happens to be the same spectrum used to model star~3.  Line
asymmetries were quantified by computing the line bisectors
\citep[e.g.,][]{Gray:92} directly from the co-added correlation
function, which is representative of the average spectral line.  We
then calculated the velocity difference in the bisector at two
different correlation levels, which we refer to as the ``bisector
span".\footnote{The particular correlation levels used here, 0.2 and
0.4, were chosen from the properties of the correlation functions for
the observations of OGLE-TR-56. Levels below 0.2 are affected by
noise, and levels above 0.4 are not measurable for some of the weaker
exposures in which the peak of the correlation function does not reach
that value.  Our conclusions below are qualitatively unaffected by
chosing different levels between these two values, so long as the
simulations use the same levels.} This is analogous to the definition
employed by \cite{Queloz:01} and \cite{Santos:02}. 

\begin{figure}
\epsscale{1.0}
\vskip 0.1in
\plotone{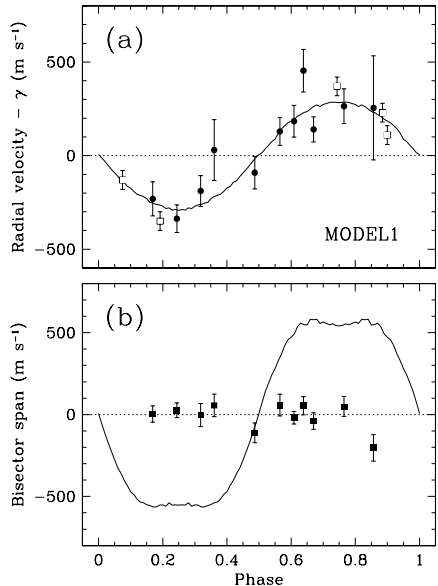}
\vskip 0.3in
 \figcaption[Torres.fig05.ps]{Simulated velocity and bisector span
variations (solid curves) from the blend scenario described by {\tt
MODEL1}, compared with the measurements of OGLE-TR-56 from
\cite{Torres:04a}. (a) Radial velocity variations, with the
center-of-mass velocity $\gamma$ subtracted. The preliminary
velocities from \cite{Mayor:03} are shown also for completeness (open
squares).  (b) Bisector span variations.\label{fig:simul_model1}}
 \end{figure}

We first simulated {\tt MODEL1}, sampling all phases of the binary
orbit and assuming $\Delta\gamma = 0$~\kms. The results are shown in
Figure~\ref{fig:simul_model1}, and are compared with the radial
velocity observations and bisector spans reported for OGLE-TR-56 by
\cite{Torres:04a}.  The top panel indicates an excellent agreement
with the measured velocities, which is however only a coincidence. The
lower panel shows that the asymmetries expected from this scenario are
much larger than observed.  The predicted bisector spans can reach
velocities exceeding $\pm$500~\ms\ near the quadratures, while the
observations show no significant variation at all within the errors.
Thus {\tt MODEL1} fails to reproduce the observations, as was already
anticipated in \S\ref{sec:lightcurve} from the excessive brightness of
star~1 relative to star~3 ($L_1/L_3 = 0.35$). 

Next we explored the sensitivity of the radial velocity and bisector
span variations to the main parameters of the blend, in an attempt to
understand their effect.  We carried out extensive simulations over a
broad range of values of $L_1/L_3$, $K_1$, $\Delta\gamma$, and $v_1
\sin i$, varying one parameter at a time and leaving the others fixed
at their nominal values for {\tt MODEL1}. A sampling of these tests is
described below. 

\begin{figure} 
 \vskip -0.1in
\includegraphics[scale=0.45]{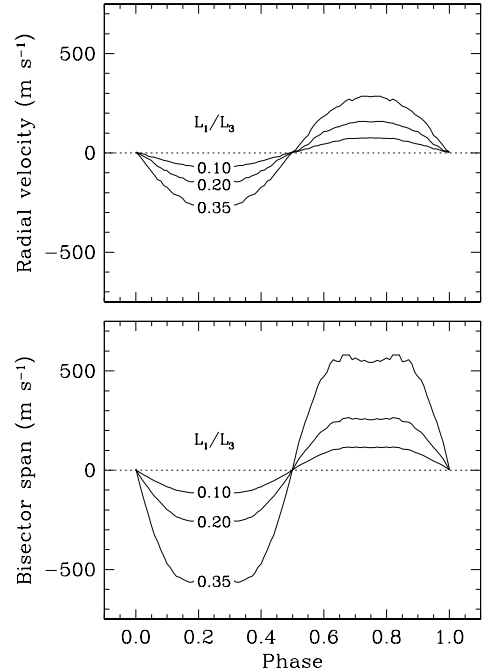} 
\vskip 0.3in
 \caption[Torres.fig06.ps]{Effect of the light ratio $L_1/L_3$ (in the
$V$ band) upon the expected radial velocity variations and bisector
spans from a blend scenario. $L_1/L_3 = 0.35$ corresponds to {\tt
MODEL1}, and the remaining parameters of the simulations are held at
their values in that model (see
Table~1).\label{fig:simul_light}}
 \end{figure}

\begin{figure}
\hskip -0.1in \includegraphics[scale=0.45]{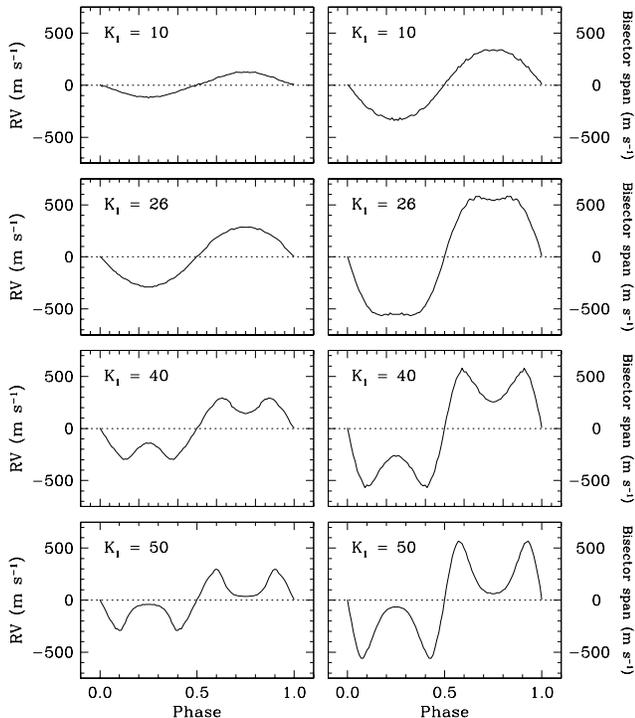}
 \vskip 0.3in
 \caption[Torres.fig07.ps]{Effect of varying the semi-amplitude $K_1$
(in~\kms) upon the expected radial velocity variations (left panels) and
bisector spans (right panels) from a blend scenario. $K_1 = 26$~\kms\
corresponds to {\tt MODEL1}, and the remaining parameters of the
simulations are held at their values in that model
(see Table~1).\label{fig:simul_kamp}}
 \end{figure}

The effect of imposing different ratios $L_1/L_3$ for the brightness
of the primary of the eclipsing binary relative to the main star in
the visual band is shown in Figure~\ref{fig:simul_light}. As expected
the amplitude of the variations scales with the brightness of star~1,
although not quite linearly. 

\begin{figure}
\vskip -0.1in
\hskip -0.1in \includegraphics[scale=0.45]{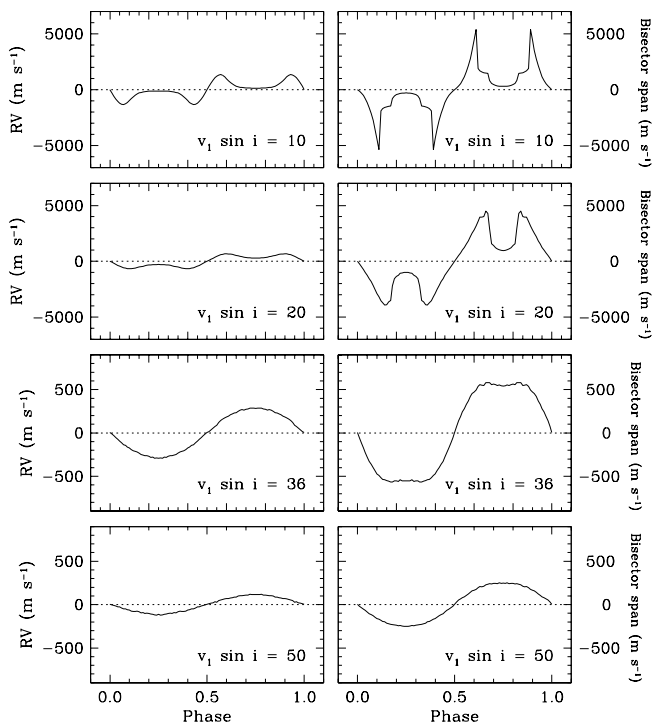}
 \vskip 0.3in
 \caption[Torres.fig08.ps]{Effect of $v_1 \sin i$ (in \kms) on the
expected radial velocity variations (left panels) and bisector spans
(right panels) from a blend scenario. Note the change in the vertical
scale for the bottom two panels on each side (see text). {\tt MODEL1}
has $v_1 \sin i = 36$~\kms.  The remaining parameters of the
simulations are held at their values in that model (see
Table~1).\label{fig:simul_vsini}}
 \end{figure}

Changing the semi-amplitude $K_1$ results in relatively minor changes
in the amplitudes, but significant changes in the shape of the
simulated curves both for the bisector span and the velocities, as
seen in Figure~\ref{fig:simul_kamp}. For small values of $K_1$ the
lines of star~1 remain within the wings of those of the main star at
all phases, and the curves are nearly sinusoidal. As $K_1$ increases,
the lines of star~1 shift out of the wings of star~3 at the
quadratures, and the effect is smaller at those phases producing a
flattening out of the curves. They become increasingly non-sinusoidal
with larger values of $K_1$.  \cite{Santos:02} described the same
effect in their simulations. 

\begin{figure}
\hskip -0.1in \includegraphics[scale=0.45]{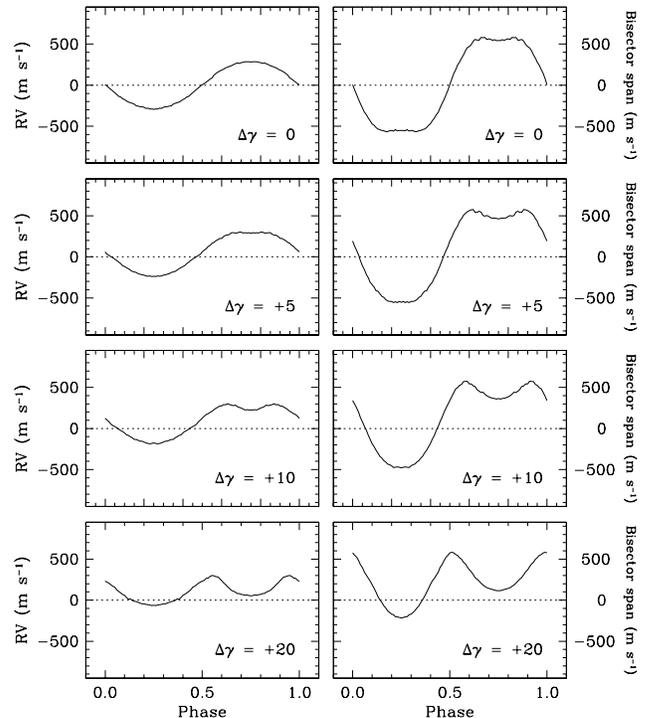}
 \vskip 0.3in
 \caption[Torres.fig09.ps]{Effect of $\Delta\gamma$ (in \kms) on the
expected radial velocity variations (left panels) and bisector spans
(right panels) from a blend scenario. {\tt MODEL1} has $\Delta\gamma =
0$~\kms. The remaining parameters of the simulations are held at their
values in that model (see
Table~1).\label{fig:simul_gamma}}
 \end{figure}

Both the amplitude and the shape of the predicted velocity variations
and bisector span variations depend very sensitively on the rotational
broadening of star~1. This is illustrated in
Figure~\ref{fig:simul_vsini}.  Sharper lines for star~1 give larger
variations except near the quadratures (for $K_1 = 26$~\kms, as in
{\tt MODEL1}), as might also be expected. This leads to cusps in the
curves that are progressively closer to phase 0.0 and 0.5 as the
rotational velocity becomes smaller, qualitatively resembling the
effect of a larger $K_1$ value.  As $v_1 \sin i$ increases, the effect
is attenuated due to the broader lines and the curves approach a more
sinusoidal shape.  In particular, a large drop in amplitude is
observed between $v_1 \sin i$ of 20~\kms\ and 30~\kms, coincident with
the point at which the rotation numerically exceeds the value of
$K_1$. 
	
Finally, a change in $\Delta\gamma$ produces another qualitative
distinction, this time an asymmetric shape to the bisector span and
velocity variations that depends upon the sign of $\Delta\gamma$.
This is shown in Figure~\ref{fig:simul_gamma}. As a result, blend
models that do not assume physical association between the eclipsing
binary and the main star (such as {\tt MODEL2}) would be expected to
show asymmetrical shapes to the bisector and velocity variations,
since $\Delta\gamma$ can be arbitrarily large in those cases.  The
measured velocity variations for OGLE-TR-56 appear to be more or less
sinusoidal (c.f.\ Figure~\ref{fig:simul_model1}), and therefore values
of $\Delta\gamma$ significantly different from zero (of order 10~\kms\
or greater, given the number of observations and their uncertainties)
would seem to be inconsistent with the observations. Similarly, values
of $K_1$ much larger than that used in {\tt MODEL1} would lead to a
shape to the radial velocity curve in disagreement with the
observations. 

\begin{figure}
\epsscale{1.2}
\vskip -0.2in
\plotone{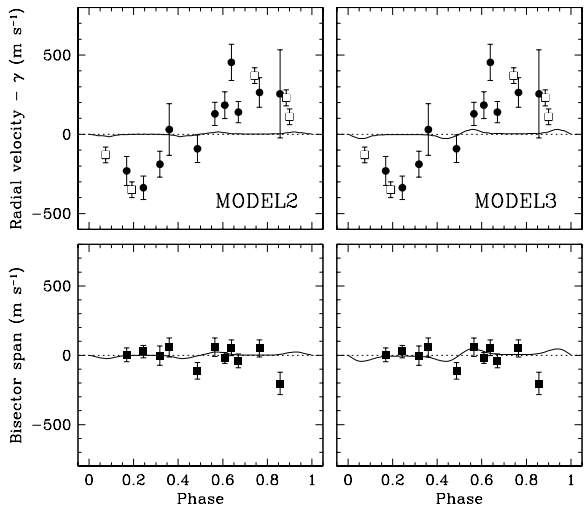}
\vskip -0.2in
 \figcaption[Torres.fig10.ps]{Simulated radial velocity and bisector
span variations (solid curves) from the blend scenarios described by
{\tt MODEL2} and {\tt MODEL3}, compared with the measurements of
OGLE-TR-56 from \cite{Torres:04a} and \cite{Mayor:03}. Details and
symbols are as in
Figure~\ref{fig:simul_model1}.\label{fig:simul_model23}}
 \end{figure}

In Figure~\ref{fig:simul_model23} we show the simulated bisector span
and velocity variations resulting from the blend scenarios described
by {\tt MODEL2} and {\tt MODEL3}, along with the observations. The
expected bisector variations from {\tt MODEL2} are less than about
25~\ms. This is actually consistent with the observations, which show
no appreciable change in the symmetry of the lines within the errors.
However, the predicted velocity variations from this blend scenario
are \emph{also} very small, with a semi-amplitude less than 15~\ms,
which is about 20 times smaller than observed. {\tt MODEL3} does not
fare much better, with a bisector semi-amplitude under 50~\ms\ and
predicted radial velocities less than $\pm$30~\ms. Thus, both of these
models clearly disagree with the observations.

\section{Discussion and concluding remarks}

The fundamental reason for the failure of all of the blend scenarios
to reproduce the observations for OGLE-TR-56 is that the variations
expected in the radial velocities and in the bisector spans from our
simulations are \emph{of the same order of magnitude}, and typically
differ by a factor of two or less, at least for configurations that
are consistent with the light curve constraints.  Similar results were
reported by \cite{Santos:02}. In our case the predicted bisector span
variations tend to be larger than the velocity variations.  By
contrast, the bisector variations actually measured for OGLE-TR-56 are
essentially negligible while the velocity variations are highly
significant. Thus, no model can reproduce both simultaneously, and our
tests show this discrepancy cannot be solved by adjusting the blend
parameters.  A further reason blend models fail here is the strong
observational constraint on the brightness of star~1. A star with only
$\sim$3\% of the light of star~3 does not produce enough of an effect
to be noticeable, either in the line profiles or in the measured
radial velocities.  The strong rotational broadening of star~1, as
predicted from our light curve fits, only reduces the effect even
further.  The conclusion must be, then, that a blend configuration
involving an eclipsing binary along the same line of sight (whether in
a physical triple system, or in the background) cannot account for the
observational signatures in OGLE-TR-56 (photometric and
spectroscopic), and we are left with a planetary companion in orbit as
the only viable explanation.  We are unable to devise a more contrived
scenario, perhaps involving more stars, that does not go against one
or another of the observational constraints. 

Dozens of transit candidates are being produced by current deep
narrow-field searches that focus on very faint stars, where crowding
is potentially a serious issue due to the high stellar densities.
Wide-field searches of brighter objects have also entered production
mode, and suffer from a similar confusion problem because of the large
angular pixel scales. The incidence of blends with a background
eclipsing binary in both types of surveys is therefore expected to be
rather high.  The same is true in most cases for future transit
searches from space such as NASA's {\it Kepler\/} mission, which will
look for even smaller photometric signals produced by Earth-size
planets. 

The procedures developed in Paper~I to model blend light curves in
detail, along with those described in \S\ref{sec:asymmetries} to
simulate spectral line asymmetries and velocity variations, are
powerful analysis tools that can be of great help for validating these
transiting planet candidates. The combination of the two techniques
makes it possible to construct self-consistent and realistic blend
scenarios and to make predictions that can then be tested
quantitatively against the observations. The application to OGLE-TR-56
in this paper has provided crucial evidence of its true nature, and
serves as an example of follow-up studies that may be applied to other
transit candidates. 
	
\acknowledgements

We are grateful to A.\ Udalski and the OGLE team for numerous
contributions to this project, and to the referee, G.\
Mall\'en-Ornelas, for a prompt and careful reading of the manuscript
and a number of helpful suggestions.  Some of the data discussed
herein were obtained at the W.\ M.\ Keck Observatory, which is
operated as a scientific partnership among the California Institute of
Technology, the University of California and the National Aeronautics
and Space Administration. The Observatory was made possible by the
generous financial support of the W.\ M.\ Keck Foundation.  G.T.\
acknowledges support for this work from NASA's {\it Kepler\/} mission,
STScI program GO-9805.02-A, the Keck PI Data Analysis Fund (JPL
1257943), and NASA Origins grant NNG04LG89G.  M.K.\ gratefully
acknowledges the support of NASA through the Michelson fellowship
program.  S.J.\ thanks the Miller Institute for Basic Research in
Science at UC Berkeley for support through a research fellowship. This
research has made use of NASA's Astrophysics Data System Abstract
Service.

\end{document}